\documentclass[english,two column]{revtex4-1}
\usepackage[LGR,T1]{fontenc}
\usepackage[latin9]{inputenc}
\usepackage{amsmath}
\usepackage{amssymb}
\usepackage{graphicx}
\usepackage{hyperref}
\hypersetup{breaklinks,
colorlinks=true,%
citecolor=blue,%
filecolor=black,%
linkcolor=blue,%
urlcolor=blue}

\makeatletter

\ProvideTextCommand{\~}{LGR}[1]{\char126#1}

\makeatother

\usepackage{babel}
\begin{document}
\title{Anisotropic dynamic excitations in a two-dimensional Fulde-Ferrell superfluid}
\author{Jinrui Ru$^{1}$}
\author{Yanxiang Zhu$^{1}$}
\author{Shuning Tan$^{2}$}
\email{sntan@ysu.edu.cn}
\author{Huaisong Zhao$^{1}$}
\email{hszhao@qdu.edu.cn}
\affiliation{$^{1}$Centre for Theoretical and Computational Physics, College of Physics, Qingdao University, Qingdao 266071, P. R. China}
\affiliation{$^{2}$Key Laboratory for Microstructural Material Physics of Hebei Province, School of Science, Yanshan University, Qinhuangdao 066004, P. R. China.}
\begin{abstract}
By calculating the dynamical structure factor of a two-dimensional (2D) Fulde-Ferrell superfluid system, the anisotropic dynamical excitations are studied systematically using random phase approximation (RPA). Our calculation results not only establish the interaction strength and the Zeeman field dependencies of the phase diagram, but also reveal the evolution of the collective modes and the single-particle excitations during the phase transition from the Bardeen-Cooper-Schrieffer (BCS) superfluid to the FF superfluid, particularly their competition with each other. The calculation results demonstrate that the optimal combination of two parameters (the interaction strength and Zeeman field) exists for finding an FF superfluid. With the increase of the angle between the transferred momentum and the center-of-mass (COM) momentum, the collective phonon mode exhibits a sharp resonance signature at small angles, which gradually diminishes as it merges into the single-particle excitations, then reappears at large angles. In an FF superfluid, the sound speed along the COM momentum direction increases with the Zeeman field strength while decreases with the interaction strength, displaying contrasting behavior compared to a BCS superfluid whereas the sound speed remains nearly constant. Notably, a remarkable roton-like dispersion emerges along the COM momentum direction while it is absent in the opposite direction. These theoretical predictions provide crucial guidance for the experimental search and study of an FF superfluid.
\end{abstract}
\maketitle

\section{Introduction}
 In the conventional Bardeen-Cooper-Schrieffer (BCS) superfluid/superconducting state, Cooper pairs carry zero center-of-mass (COM) momentum. Interestingly, there may be a type of exotic superfluid with non-zero COM momentum ${\bf Q}$ in a polarized Fermi system, known as the Fulde-Ferrell-Larkin-Ovchinnikov (FFLO) superfluid \cite{Kinnunen2018}. The order parameter in an FFLO superfluid exhibits spatial dependence and manifests in two forms: the plane wave form $\Delta({\bf r})=\Delta_{0}e^{i{\bf Q \cdot r}}$ for the Fulde-Ferrell (FF) superfluid \cite{Fulde1964}, and the standing wave form $\Delta({\bf r})=\Delta_{0}\cos[{\bf Q \cdot r}]$ for the Larkin-Ovchinnikov (LO) type \cite{Larkin1965,Huang2022}. Recent experimental achievement of a LO-type pair-density wave (PDW) in condensed matter physics \cite{Agterberg2020,Chen2021,Liu2023,Kittaka2023,Kasahara2020} has sparked interest in realizing an FFLO superfluid state in ultracold atomic gases \cite{Liu2007,Huhui2018,xu2014,Liuxj2013,Zhang2013,Cao2014,Kawamura2022,Pini2021}.
 Compared with the BCS/LO superfluid, theoretically the FF state has a substantially narrower parameter space and exhibits smaller pairing gaps, posing significant challenges to its experimental implementation \cite{Liu2007,Sheehy2007,Akbar2016,Xub2014}.
 Both the interaction strength and the Zeeman field can modulate the pairing gap magnitude, the optimal parameter regime for finding an FF superfluid state in  polarized Fermi gases remains unresolved. A fundamental question is raised: How can an FF superfluid be detected and distinguished from the BCS superfluid?

 In a neutral superfluid, the breaking of $U(1)$ symmetry gives rise to a low-energy collective mode. This mode corresponds to a collective movement of all particles throughout the system and is identified as the gapless Goldstone mode \cite{Nambu1960,Goldstone1962}.
Dynamical structure factor is often employed to investigate these dynamical excitations \cite{Combescot06,Combescot2006,Zou2021,Watabe2010}. As a two-body correlation physical observable originated from the density-density correlation functions, the dynamical structure factor reveals essential many-body information about the system.
Generally, the collective modes appear in a small transferred momentum region and the single-particle excitations are mainly studied in a large transferred momentum region \cite{Zhao2020}. In particular, a two-photon Bragg scattering technique has been developed to measure the dynamical structure factor from one-dimensional (1D) to three-dimensional (3D) Fermi gases \cite{Pagano2014,Sobirey2022,Veeravalli08,Hoinka17,Biss2022,Senaratne2022,Li2022,Dyke2023}.
Numerical simulations using the quantum Monte Carlo (QMC) methods have successfully reproduced the dynamical structure factor in Fermi superfluids  \cite{Vitali2020,Vitali2022,Apostoli2024}. Theoretically, some works have studied the dynamical excitations from the conventional BCS superfluids to the topological superfluids in low-dimensional systems \cite{Jayantha2012,Koinov2017,Zhao2023,Gao2023,Zhao2024}. Our previous study demonstrates significant differences in dynamical excitations between BCS and topological superfluid phases. Notably, under the Zeeman field variations, the sound speed remains nearly constant in a BCS superfluid but exhibits abrupt enhancement when entering the topological superfluid \cite{Gao2023,Zhao2024}. These distinct signatures of the dynamical excitations may provide an effective approach for identifying the FF superfluid.

In an FF superfluid with fixed COM momentum ${\bf Q}$, the band structure exhibits asymmetry relative to the ${\bf Q}$ direction \cite{Kinnunen2018,Liu2007,Suba2020,Wu2013,Zhou2013}, leading to the anisotropic collective modes \cite{Samokhin2011,Edge2009,Edge2010,Heikkinen2011,Koinov2011,Boyack2017}. Numerical simulations using random phase approximation (RPA) theory investigate the collective modes in a 2D optical lattice with 40,000 lattice sites \cite{Heikkinen2011}. Their results revealed that the anisotropic pairing in FFLO states produces an anisotropic sound speed, where the sound speed parallel to ${\bf Q}$ is greater than that perpendicular to ${\bf Q}$.
In addition to the low-energy Goldstone phonon mode, Koinov {\it et al.} identified two asymmetric roton-like collective modes in an FF superfluid state of 2D polarized optical lattice \cite{Koinov2011}. The existing papers primarily focus on the collective modes on optical lattices and lack systematic investigation of the dynamical excitations, particularly regarding the competition between the collective modes and the single-particle excitations. In addition to the Cooper pairs, many unpaired atoms of an FF superfluid at the Fermi surface give rise to the gapless single-particle excitations. Thus the competition between the single-particle excitations and the collective modes is intense. In this paper, we theoretically investigate dynamical excitations in an FF superfluid state of 2D continuous Fermi gases and analyze the asymmetric characteristics of both the collective modes and the single-particle excitations. We also provide the optimal parameters for FF superfluid formation and indicate that excessive interaction strength hinders this state's realization.

This paper is organized as follows. In Sec \ref{modelH}, we derive the equations of motion of the Green's functions for 2D polarized Fermi gases within the mean-field approximation. In Sec. \ref{phased}, we discuss the ground state under the Zeeman field, present the phase diagram, and analyze anisotropic quasiparticle spectra.
In Sec. \ref{RPADSF}, we derive the dynamical structure factor using the RPA theory.  We present results of dynamic structure factor and discuss the anisotropic sound speed and complex single-excitations in an FF superfluid in Sec. \ref{FFDSF}. In Sec. \ref{discussion}, we provide a discussion. Finally, we give our conclusions in Sec. \ref{summary}. Some calculation details are provided in the Appendix.

\section{Model and Hamiltonian}
\label{modelH}
 For a polarized two-component Fermi superfluid with a s-wave contact interaction, the Hamiltonian can be described in momentum space as follows \cite{Kinnunen2018,Zoup2024}:
\begin{eqnarray}\label{FFHmodel}
H=\sum_{{\bf k},\sigma} (\varepsilon_{\bf k} - \mu_\sigma) \psi_{{\bf k}\sigma}^{\dagger}\psi_{{\bf k}\sigma}+U\sum_{\bf k}\psi_{{\bf k}\uparrow}^{\dagger}\psi_{{\bf Q-k}\downarrow}^{\dagger}\psi_{{\bf Q-k}\downarrow}\psi_{{\bf k}\uparrow},\nonumber\\
\end{eqnarray}
where $\varepsilon_{\bf k}={\bf k}^{2}/(2m)$, $\psi_{{\bf k}\sigma}$ ($\psi^{\dagger}_{{\bf k}\sigma}$) is the annihilation (generation) operator for spin-$\sigma$ component with momentum {\bf k}, $\sigma=\uparrow$ or $\downarrow$. $\mu_{\sigma}$ is the chemical potential, and $U$ is the 2D bare interatomic attractive interaction strength. Here and thereafter, we always set $\hbar=k_{B}=1$. The Fermi wave vector ${k_{F}=\sqrt{2{\pi}n}}$ (noninteracting Fermi system) and Fermi energy ${\varepsilon_{F}={k^{2}_{F}}/(2m)}$ are used as momentum and energy
units, respectively. In an FF superfluid state, we define the order parameter, namely, the paring gap $\Delta^{*}=U<\psi_{{\bf k}\uparrow}^{\dagger}\psi_{{\bf Q-k}\downarrow}^{\dagger}>$. Within the mean-field approximation, the four-operator term in Eq. (\ref{FFHmodel}) can be dealt into a two-operators one as,
$U\psi_{{\bf k}\uparrow}^{\dagger}\psi_{{\bf Q-k}\downarrow}^{\dagger}\psi_{{\bf Q-k}\downarrow}\psi_{{\bf k}\uparrow}=\Delta^{*}\psi_{{\bf Q-k}\downarrow}\psi_{{\bf k}\uparrow}+\Delta\psi_{{\bf k}\uparrow}^{\dagger}\psi_{{\bf Q-k}\downarrow}^{\dagger}-\Delta^{2}/U$. Generally, we define the mean chemical potential $\mu=(\mu_{\uparrow}+\mu_{\downarrow})/2$ and the effective Zeeman field $h=(\mu_{\uparrow}-\mu_{\downarrow})/2$ in a polarized Fermi gas.
 Then the above Hamiltonian is displayed into a mean-field one as
\begin{eqnarray}\label{mean-field}
H_{MF}&=& \sum_{{\bf k},\sigma} (\xi_{\bf k} - h\sigma_{z}) \psi_{{\bf k}\sigma}^{\dagger}\psi_{{\bf k}\sigma}\nonumber\\
&+&\sum_{{\bf k}} (\Delta\psi_{{\bf k}\uparrow}^{\dagger}\psi_{{\bf Q-k}\downarrow}^{\dagger}+\Delta^{*} \psi_{{\bf Q-k}\downarrow}\psi_{{\bf k}\uparrow})-\frac{\Delta^2}{U},
\end{eqnarray}
where $\xi_{\bf k}=\varepsilon_{\bf k}-\mu$.

The above mean-field Hamiltonian can be solved by the equations of motion of the Green's functions. We define the spin-up Green's function, $G_{\uparrow}({\bf k},\tau-\tau{'})=-\left\langle T_{\tau} \psi_{{\bf k}\uparrow}(\tau)\psi^{\dagger}_{{\bf k}\uparrow}(\tau{'})\right\rangle$, the spin-down Green's function, $G_{\downarrow}({\bf k},\tau-\tau{'})=-\left\langle T_{\tau} \psi_{{\bf k}\downarrow}(\tau)\psi^{\dagger}_{{\bf k}\downarrow}(\tau{'})\right\rangle$, and the singlet pairing one $\Gamma^{\dagger}({\bf k},\tau-\tau{'})=-\left\langle T_{\tau} \psi^{\dagger}_{{\bf Q-k}\downarrow}(\tau)\psi^{\dagger}_{{\bf k}\uparrow}(\tau{'})\right\rangle$, respectively. The Green's functions $G_{\uparrow}$, $G_{\downarrow}$ correspond to the normal particle density and the $\Gamma^{\dagger}$ is related to the singlet Cooper pairing information. Owing to the appearance of the Zeeman field, $G_{\uparrow}\neq G_{\downarrow}$. The expressions of these three Green's functions are obtained in a BCS form by
\begin{subequations}\label{bcsgreen}
 \begin{eqnarray}
G_{\uparrow}\left({\bf k},\omega\right)&=&\frac{U_{\bf k}^2}{\omega-E^{(1)}_{\bf k}}+\frac{V_{\bf k}^2}{\omega+E^{(2)}_{\bf k}}\\
G_{\downarrow}\left({\bf k},\omega\right)&=&\frac{U_{\bf k}^2}{\omega-E^{(2)}_{\bf Q-k}}+\frac{V_{\bf k}^2}{\omega+E^{(1)}_{\bf Q-k}}\\
\Gamma^{\dagger}\left(\bf{k},\omega\right)&=&\frac{\Delta^{*}}{2E_{\bf{k}}}
\left(\frac{1}{\omega-E^{(1)}_{\bf k}}-\frac{1}{\omega+E^{(2)}_{\bf k}}\right),
\end{eqnarray}
\end{subequations}
where $U_{\bf k}^2=0.5[1+0.5(\xi_{\bf k}+\xi_{\bf Q-k})/E_{\bf{k}}]$, $V_{\bf k}^2=0.5[1-0.5(\xi_{\bf k}+\xi_{\bf Q-k})/E_{\bf{k}}]$. $E^{(1)}_{\bf k}$, $E^{(2)}_{\bf k}$ are the quasiparticle spectra and can be obtained as,
\begin{eqnarray}\label{Quasi-en-spec}
E^{(1)}_{\bf k}&=&E_{\bf{k}}+\frac{\xi_{\bf k}-\xi_{\bf Q-k}}{2}-h,\nonumber\\
E^{(2)}_{\bf k}&=&E_{\bf{k}}-\frac{\xi_{\bf k}-\xi_{\bf Q-k}}{2}+h.
\end{eqnarray}
Here $E_{\bf{k}}=\sqrt{{(\xi_{\bf k}+\xi_{\bf Q-k})}^2/4+{\Delta^{2}}}$. The value of $-E^{(2)}_{\bf k}$ is always negative relative to the Fermi energy, while $E^{(1)}_{\bf k}$ passes through the Fermi energy.

 The thermodynamic potential of a system $\Omega=-\ln{Z}/\beta$, where $\beta=1/T$, $T$ is the temperature, $Z={\rm T_{r}}[e^{-\beta{H}}]$ is the partition function, ${\rm T_{r}}$ is the trace. Based on the Eq. (\ref{mean-field}), the mean-field thermodynamic potential is given by,
\begin{eqnarray}\label{thermalpotential}
\Omega&=&\sum_{\bf k}\left(\frac{\xi_{\bf k}+\xi_{\bf Q-k}}{2}-E_{\bf k}\right)-\frac{\Delta^2}{U}\nonumber\\
&-&T\sum_{\bf k}\ln{(1+e^{-\beta E^{(1)}_{\bf k}})(1+e^{-\beta E^{(2)}_{\bf k}})}.
\end{eqnarray}
The average chemical potential $\mu$, the pairing gap $\Delta$, and the COM momentum $Q$ are determined by using the self-consistent
 stationary conditions, namely, $N=-\partial{\Omega}/\partial{\mu}$, $\partial{\Omega}/\partial{\Delta}=0$, $\partial{\Omega}/\partial{Q}=0$, respectively. These three self-consistent equations can be obtained by
\begin{subequations}\label{threeequtions}
\begin{eqnarray}
N&=&\sum_{\bf k}\left[1-D({\bf k},T)\frac{\xi_{\bf k}+\xi_{\bf Q-k}}{2E_{\bf k}}\right],\\
-\frac{1}{U} &=& \sum_{\bf k}\frac{D({\bf k},T)}{2E_{\bf k}},\\
0&=&\sum_{\bf k}D({\bf k},T)(1-\frac{\xi_{\bf k}+\xi_{\bf Q-k}}{2E_{\bf k}})\frac{\partial{\xi_{\bf Q-k}}}{\partial{Q}},
\end{eqnarray}
\end{subequations}
where the function $D({\bf k},T)=1-n_{F}(E^{(1)}_{\bf k})-n_{F}(E^{(2)}_{\bf k})$, $n_{F}(x)=1/(e^{\beta{x}}+1)$ is the Fermi distribution. $\partial{\xi_{\bf Q-k}}/\partial{Q}=2(Q-k\cos{\varphi})$, and $\varphi$ is the angle between ${\bf k}$ and ${\bf Q}$ and is integrated during the summation.  To eliminate the divergence of Eq. \ref{threeequtions}(b) introduced by an s-wave contact interaction, the bare interaction strength $U$ should be regularized by $1/U=-\sum_{\bf k}1/({2\varepsilon_{\bf k}+E_{\rm b}})$, where $E_{\rm b}$ is the magnitude of the binding energy. Thus, the pairing gap equation can be rewritten as,
\begin{eqnarray}\label{regulareq}
\sum_{\bf k}\left(\frac{D({\bf k},T)}{2E_{\bf k}}-\frac{1}{2\varepsilon_{\bf k}+E_{\rm b}}\right)=0.
\end{eqnarray}
Under the given $h$, $T$ and $E_{\rm b}$, the value of $\mu$, $\Delta$ and $Q$ can be solved self-consistently.
We consider a typical low temperature $T=0.01{\varepsilon_{F}}$ (close to zero) to avoid numerical divergence induced by zeros of $E^{(1)}_{\bf k}$. We adopt the {\it plane polar coordinate system} with $Q$ aligned along the polar axis and $\varphi$ as the polar angle.
\section{Phase diagram and anisotropic quasiparticle spectra}
\label{phased}
\begin{figure}[h!]
\includegraphics[scale=0.45]{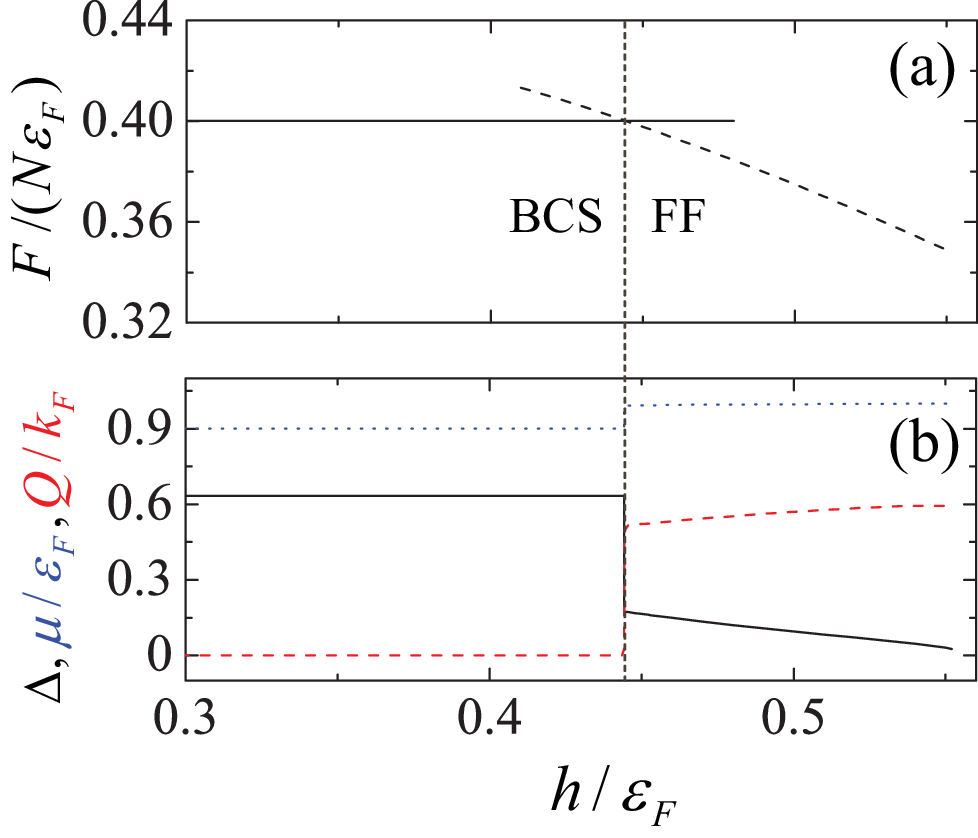}
\caption{(a) Free energy $F=\Omega+{\mu}N$, (b) chemical potential $\mu$ (blue dotted line), pairing gap $\Delta$ (black solid line), and COM momentum $Q$ (red dashed line) as a function of the Zeeman field $h$ for $E_{\rm b}=0.2{\varepsilon_{F}}$, respectively. The vertical dotted line marks the BCS-FF superfluid phase transition at $h_{\rm c}=0.444{\varepsilon_{F}}$.
\label{fig1}}
\end{figure}

In polarized Fermi superfluid gases, a phase transition from a conventional BCS-type superfluid to an FF-type superfluid state occurs when the critical Zeeman field strength $h_{\rm c}$ is reached. The ground state of the system is determined by minimizing the free energy, $F=\Omega+{\mu}N$.  We therefore plot $F$ as a function of $h$ in Fig. \ref{fig1}\textcolor{blue}{(a)}. The corresponding chemical potential $\mu$, pairing gap $\Delta$ and the COM momentum $Q$ are shown in Fig. \ref{fig1}\textcolor{blue}{(b)}.
In the low$-h$ region, the ground state corresponds to a conventional BCS superfluid with $Q=0$. Here, $\mu$ and $\Delta$ remain nearly constant as $h$ increases, consistent with the Meissner effect. For $h>h_{\rm c}$, the FF superfluid with non-zero $Q$ exhibits the lower free energy than the BCS superfluid, indicating that a first-order phase transition happens. The FF superfluid thus becomes the new ground state and exhibits the coexistence of pairing and non-zero magnetization. Both $\mu$ and $\Delta$ show abrupt changes at $h_{\rm c}$ and vary with $h$. Near $h_{\rm c}=0.444{\varepsilon_{F}}$, the pairing gap transitions from $\Delta_{\rm BCS}=0.6325{\varepsilon_{F}}$ in the BCS state to $\Delta_{\rm FF}=0.1747{\varepsilon_{F}}$ with ${Q}=0.5135{k_{F}}$ in the FF side.  In an FF superfluid, pairing occurs only on one side of each Fermi surface, resulting in fewer atoms participating in Cooper pairing. Thus the pairing gap becomes small when entering an FF superfluid. In contrast, the LO superfluid exhibits much larger pairing gap \cite{Loh2010}. Because the pairing occurs on both sides of each Fermi surface, which enhancing the number of atoms participating in Cooper pairing.

\begin{figure}[h!]
\includegraphics[scale=0.35]{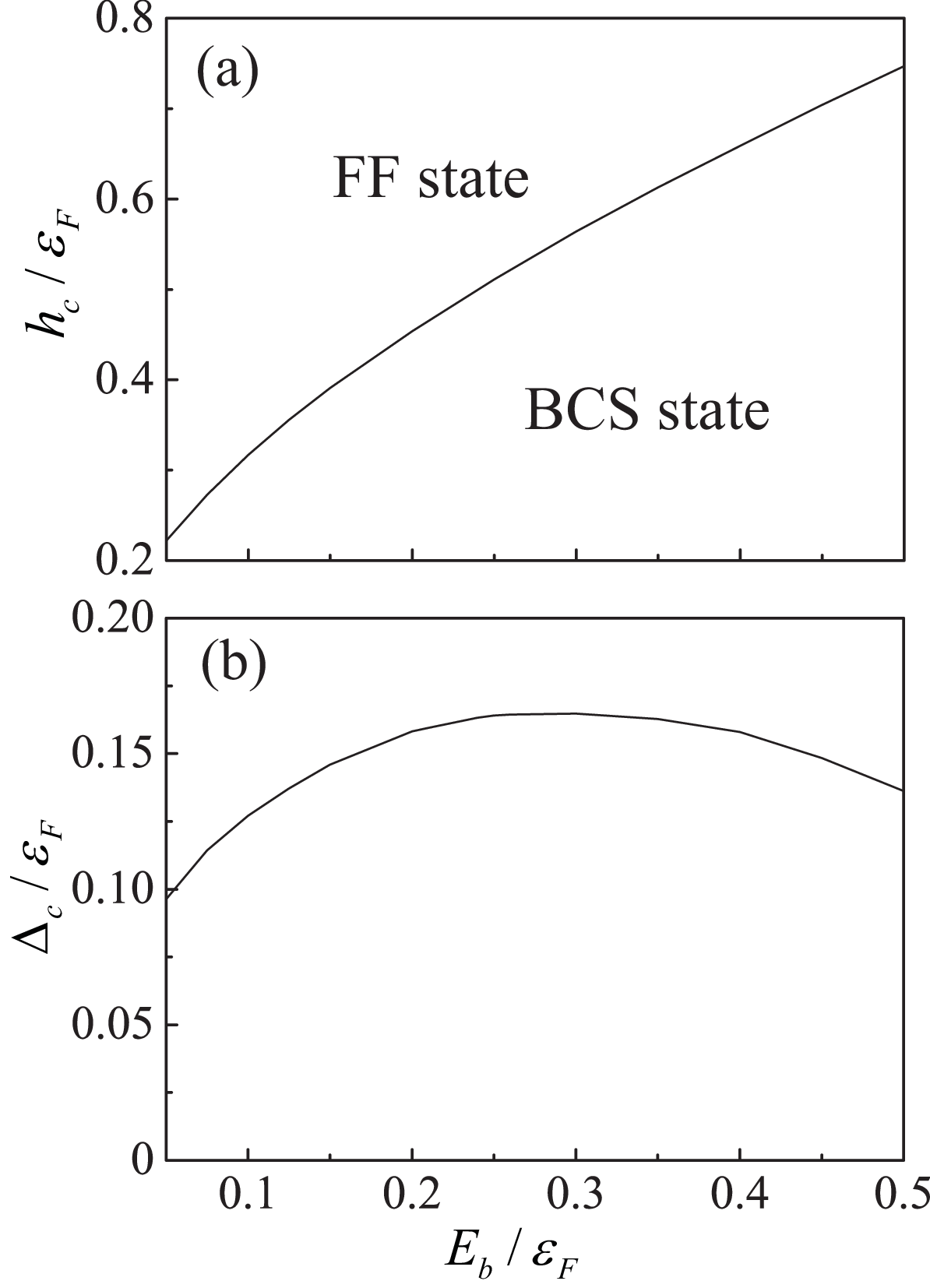}
\caption{ (a) Critical Zeeman field $h_{\rm c}$ and (b) its related pairing gap $\Delta_{\rm c}$ at different interaction strength $E_{\rm b}$. And $h_{\rm c}$ marks the BCS-FF superfluid phase transition.\label{fig2}}
\end{figure}

The physical parameters at the phase transition point are crucial for experimental detection of an FF superfluid, particularly the pairing gap. An important question arises: Does stronger interaction strength facilitate the formation of an FF superfluid? We demonstrate that $h_{\rm c}$ depends on the interaction strength. Figs. \ref{fig2}\textcolor{blue}{(a)} and \textcolor{blue}{(b)} show $h_{\rm c}$ and its corresponding pairing gap $\Delta_{\rm c}$ as a function of interaction strength $E_{b}$, respectively.
 Obviously, $h_{\rm c}$ is proportional to $E_{\rm b}$, indicating that under stronger interaction strength, a higher $h_{\rm c}$ is required during the BCS-FF superfluid phase transition. However, the stronger interaction strength does not guarantee that it is easier to detect an FF superfluid. From Fig. \ref{fig1}, the FF superfluid's pairing gap $\Delta_{\rm FF}$ of is significantly smaller than that in the BCS state $\Delta_{\rm BCS}$, leading to a lower critical temperature. This small pairing gap poses major experimental challenges for search and study of an FF superfluid. Therefore, searching for the larger $\Delta_{\rm FF}$ is the key. From Fig. \ref{fig2}\textcolor{blue}{(b)}, the pairing gap at the critical point $\Delta_{\rm c}$ shows a dome-shaped dependence on $E_{\rm b}$: $\Delta_{\rm c}$ initially grows with $E_{\rm b}$, and reaches a maximum at $E_{\rm b}\approx0.3{\varepsilon_{F}}$, then decreases when $E_{\rm b}>0.3{\varepsilon_{F}}$. Under the strong attraction interaction ($E_{\rm b}\gg0.3{\varepsilon_{F}}$), the spin-up and spin-down atoms form a condensate of tightly bounded Cooper pairs, suppressing the distortion of Fermi surfaces (the hallmark of an FF superfluid). Therefore, to realize an FF superfluid experimentally, it is necessary to optimize both $E_{\rm b}$ and $h$. With further increase in $h$ beyond $h_{\rm c}$, $\Delta_{\rm FF}$ monotonically decreases until $\Delta_{\rm FF}=0$, which marks a second-order phase transition from an FF superfluid to the polarized normal-state Fermi gases.

\begin{figure}[h!]
\includegraphics[scale=0.45]{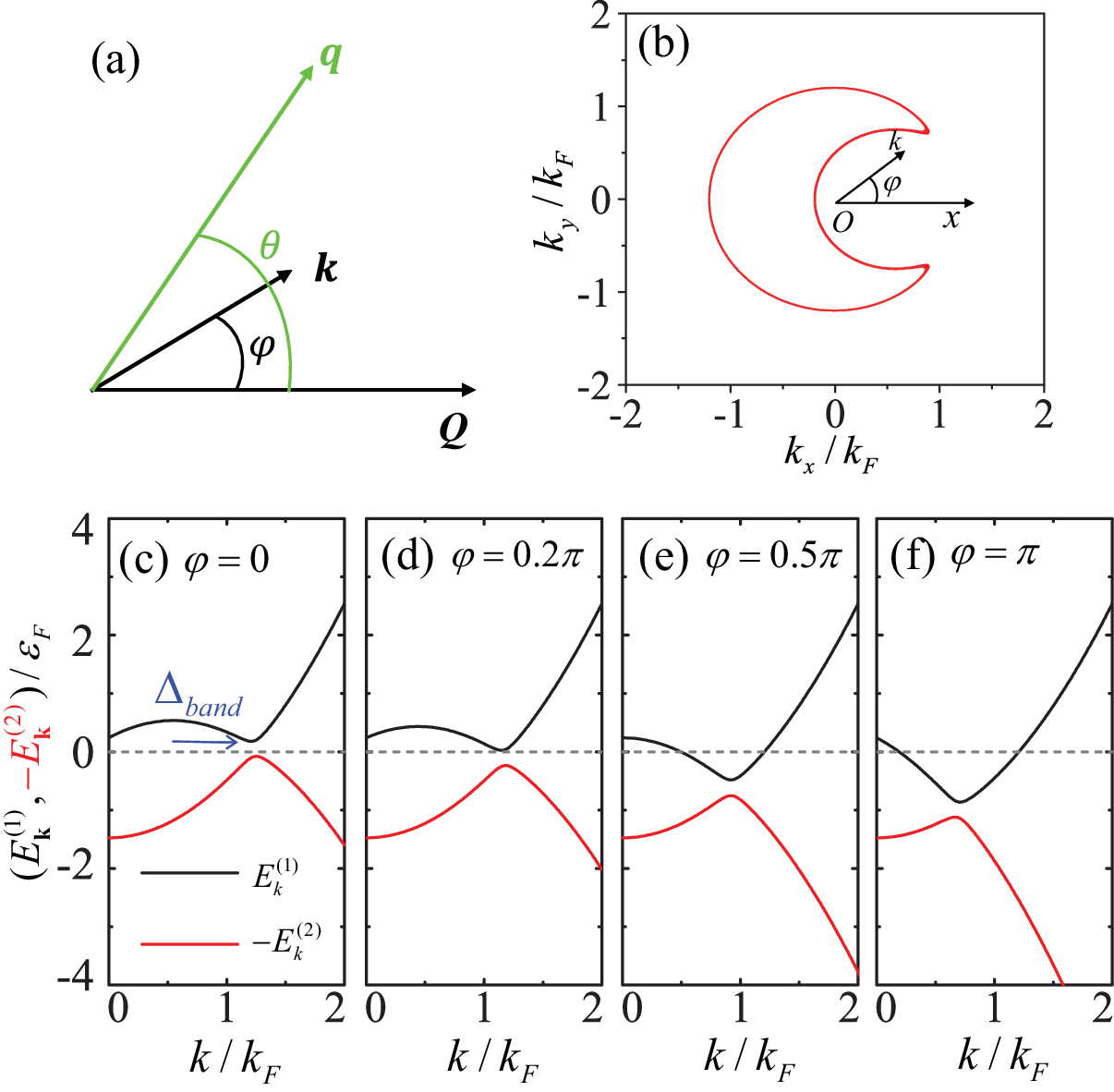}
\caption{(a) Momentum vectors ${\bf k}$, ${\bf Q}$ (COM momentum), and the angles $\varphi$ (between ${\bf k}$ and ${\bf Q}$), $\theta$ (between the transferred momentum ${\bf q}$ and ${\bf Q}$); (b) the Fermi surface for $E^{(1)}_{\bf k}=0$; (c)-(f) quasi-particle spectra $E^{(1)}_{\bf k}$ (black line), $-E^{(2)}_{\bf k}$ (red line) as a function of ${\bf k}$ for $\varphi=0, 0.2\pi, 0.5\pi,\pi$. Parameters: $E_{\rm b}=0.2{\varepsilon_{F}}$, $h=0.47{\varepsilon_{F}}$, $Q=0.543k_{F}$. The blue arrow marks the band gap $\Delta_{\rm band}$ between $E^{(1)}_{\bf k}$ and ${\varepsilon_{F}}$.
\label{fig3}}
\end{figure}

 Many physical properties of an FF superfluid exhibit anisotropy owing to the COM momentum ${\bf Q}$, including the quasi-particle spectra and the dynamical excitations. We now study the anisotropy of the quasiparticle spectra of an FF superfluid in Fig. \ref{fig3}. The fixed ${\bf Q}$ direction defines $\varphi$ as the angle between a certain momentum ${\bf k}$ and ${\bf Q}$, illustrated in Fig. \ref{fig3}\textcolor{blue}{(a)}. Since $E^{(1)}_{\bf k}$ crosses the Fermi energy, we plot its corresponding Fermi surface as a function of ${\bf k}$ in Fig. \ref{fig3}\textcolor{blue}{(b)}.
Our results reveal the angular anisotropy in both $E^{(1)}_{\bf k}$ and $E^{(2)}_{\bf k}$. To see this point clearly, we respectively calculate the
momentum dependence of $E^{(1)}_{\bf k}$ and $E^{(2)}_{\bf k}$ at fixed angles $\varphi=0, 0.2\pi, 0.5\pi,\pi$, shown in Figs. \ref{fig3}\textcolor{blue}{(c)-(f)}. It is shown that $E^{(1)}_{\bf k}$ and $E^{(2)}_{\bf k}$ as a function of $\varphi$ show a different momentum behavior: at $\varphi=0$, a finite band gap $\Delta_{\rm band}$ exists between $E^{(1)}_{\bf k}$ and ${\varepsilon_{F}}$; $\Delta_{\rm band}$ closes gradually for $\varphi=0.2\pi$; as $\varphi$ further increases, $E^{(1)}_{\bf k}$ passes through ${\varepsilon_{F}}$. This angular dependence behavior originates from Eq. (\ref{Quasi-en-spec}), $E^{(1)}_{\bf k}=\sqrt{{(\xi_{\bf k}+\xi_{\bf Q-k})}^2/4+{\Delta^{2}}}+({\xi_{\bf k}-\xi_{\bf Q-k}})/{2}-h$, where $\xi_{\bf Q-k}={\bf (Q- k)}^{2}-\mu=Q^{2}-2kQ\cos{\varphi}+k^{2}-\mu$.
\section{Dynamical structure factor and random phase approximation}
\label{RPADSF}
The mean-field theory neglects the contribution from the fluctuation term in the interaction Hamiltonian, and therefore can not reliably predict the dynamical excitations in interacting systems, especially for the collective modes. To account for fluctuations \cite{Liu2004,He2016,Ganesh2009}, the random phase approximation (RPA) has proven effective for calculating response function beyond the mean-field level. RPA theory has been widely used to investigate the collective modes. For example, in a 3D Fermi superfluid, the dynamical excitations calculated through RPA theory even show quantitative agreement with the experimental results \citep{Biss2022,Zou2018,Zou2010}. The 2D theoretical predictions using RPA qualitatively agree with the QMC data \cite{Zhao2020,Zhao2023-3}. Thus, it is reasonable to expect that this RPA strategy should provide qualitatively reliable predictions for a 2D FF superfluid.

We briefly shed light on the main idea of RPA theory for investigating dynamical excitations. As a beyond mean-field strategy, RPA theory incorporates fluctuations in the Hamiltonian. In an FF superfluid, there are four different density operators, with the normal spin-up/down densities $\hat{n}_{\uparrow}=\psi^{\dagger}_{\uparrow}\psi_{\uparrow}$, $\hat{n}_{\downarrow}=\psi^{\dagger}_{\downarrow}\psi_{\downarrow}$, and the anomalous pairing operator and its complex conjugate $\hat{n}_{\Delta}=\psi_{\downarrow}\psi_{\uparrow}$, $\hat{n}_{\Delta^{*}}=\psi^{\dagger}_{\uparrow}\psi^{\dagger}_{\downarrow}$ (describe the Cooper pairs). These four densities are coupled by atomic interactions. Any perturbations in one density induce fluctuations in others. Within the framework of linear response theory, a small external perturbation potential $V_{\rm ext}$ induces the density fluctuations $\delta n$, and the response function is defined as: $\delta n=\chi V_{\rm ext}$.

The main idea of RPA is to treat fluctuation Hamiltonian as part of an effective external potential. The response function $\chi$ beyond mean-field theory is connected to its mean-field approximation $\chi^0$ by
 \begin{eqnarray}\label{chi}
 \chi({\bf q},i\omega_{n})=\frac{\chi^{0}({\bf q},i\omega_{n})}{\hat{1}-\chi^{0}({\bf q},i\omega_{n})U{M_{I}}}.
\end{eqnarray}
Here $M_{I}=\sigma_{0}\otimes\sigma_{x}$ is a direct product of unit matrix $\sigma_{0}$ and Pauli matrix $\sigma_{x}$. The mean-field response function $\chi^0$ is easy to obtain, and its expression is a $4\times4$ matrix,
\begin{eqnarray}\label{matrix}
\chi^{0}({\bf q},i\omega_{n})=\left[
\begin{array}{cccccc}
&\chi^{0}_{{\uparrow}{\uparrow}}&\chi^{0}_{{\uparrow}{\downarrow}}&\chi^{0}_{{\uparrow}{\Delta}}&\chi^{0}_{{\uparrow}\Delta^{*}}\\
&\chi^{0}_{{\downarrow}{\uparrow}} &\chi^{0}_{{\downarrow}{\downarrow}}&\chi^{0}_{{\downarrow}{\Delta}}&\chi^{0}_{{\downarrow}\Delta^{*}}\\
&\chi^{0}_{{\Delta}{\uparrow}}&\chi^{0}_{{\Delta}{\downarrow}}&\chi^{0}_{{\Delta}{\Delta}}&\chi^{0}_{{\Delta}\Delta^{*}}\\
&\chi^{0}_{\Delta^{*}{\uparrow}}&\chi^{0}_{\Delta^{*}{\downarrow}}&\chi^{0}_{\Delta^{*}{\Delta}}&\chi^{0}_{\Delta^{*}\Delta^{*}}\\
\end{array}
\right].
 \end{eqnarray}
 The dimension of $\chi^0$ reflects the coupling situation among four density channels. These 16 matrix elements are determined by the density-density correlation functions derived from the previous defined Green's functions. For example, $\chi^{0}_{{\uparrow}{\downarrow}}=-\left\langle T_{\tau} \psi^{\dagger}_{\uparrow}({\bf r},\tau)\psi_{\uparrow}({\bf r},\tau)\psi^{\dagger}_{\downarrow}({\bf r'},\tau{'})\psi_{\downarrow}({\bf r'},\tau{'})\right\rangle$, which contracts via Wick's theorem to $\chi^{0}_{{\uparrow}{\downarrow}}=-\Gamma^{\dagger}({\bf r}-{\bf r'},\tau-\tau')\Gamma({\bf r'}-{\bf r},\tau'-\tau)$.
 Considering system symmetries, only 9 of these matrix elements are independent, i.e.,
 $\chi^{0}_{{\uparrow}{\downarrow}}=\chi^{0}_{{\downarrow}{\uparrow}}=-\chi^{0}_{{\Delta}{\Delta}}=-\chi^{0}_{\Delta^{*}\Delta^{*}}$,
 $\chi^{0}_{{\Delta}{\uparrow}}=\chi^{0}_{{\uparrow}\Delta^{*}}$, $\chi^{0}_{{\Delta}{\downarrow}}=\chi^{0}_{{\downarrow}\Delta^{*}}$
 $\chi^{0}_{\Delta^{*}{\downarrow}}=\chi^{0}_{{\downarrow}{\Delta}}$, $\chi^{0}_{\Delta^{*}{\uparrow}}=\chi^{0}_{{\uparrow}{\Delta}}$.
 Their explicit forms are provided in the Appendix.

The total density response function $\chi_n$ is defined as $\chi_n \equiv\chi_{{\uparrow}{\uparrow}}+\chi_{{\uparrow}{\downarrow}}+\chi_{{\downarrow}{\uparrow}}+\chi_{{\downarrow}{\downarrow}}$.
From the fluctuation-dissipation theory, the density dynamical structure factor $S({\bf q},{\omega})$ is obtained as,
\begin{eqnarray}\label{sqw}
  S({\bf q},{\omega})&=&-\frac{1}{\pi}\frac{1}{1-e^{-\omega/T}}{\rm Im}\chi_n\left({\bf q},i\omega_{n}\to \omega+i\delta\right),
 \end{eqnarray}
where ${\bf q}$ and $\omega$ denote the transferred momentum and energy, respectively. $\delta$ is a small positive number in numerical calculations (usually we set $\delta=0.001$).

\section{Results}
 \label{FFDSF}
\begin{figure}[h!]
\centering
\includegraphics[width=0.48\textwidth]{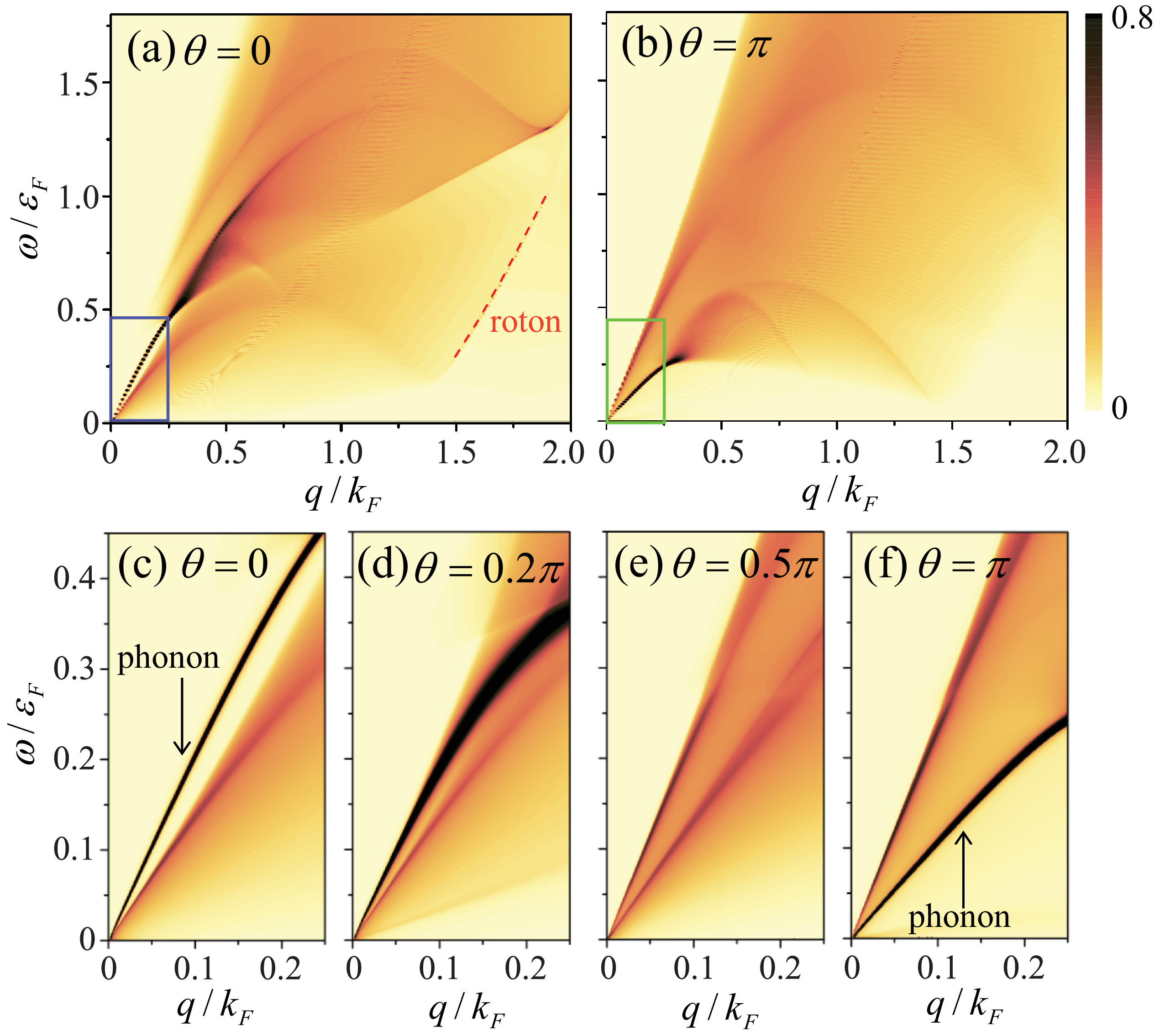}
\caption{\label{fig4} Color maps of $S({\bf q},{\omega})$ in an FF state for (a) $\theta=0$, and (b) $\theta=\pi$ from $q=0$ to $q=2.0k_{F}$ with $h=0.47{\varepsilon_{F}}$, $E_{\rm b}=0.2{\varepsilon_{F}}$. The detailed low-momentum region $(q\in[0, 0.25k_{F}])$ for (c) $\theta=0$ (the blue square region in (a)), (d) $\theta=0.2\pi$, (e) $\theta=0.5\pi$, and (f) $\theta=\pi$ (the green square region in (b)), respectively. The arrows indicate the phonon mode positions.}
\end{figure}

 We firstly discuss the dynamical structure factor $S({\bf q},{\omega})$ in an FF superfluid of 2D Fermi gases. By analyzing $S({\bf q},{\omega})$ under different transferred momenta, one can obtain both the collective excitations and the single-particle excitations. We have calculated the energy and momentum dependence of $S({\bf q},{\omega})$, and the contour plots of $S({\bf q},{\omega})$ for (a) $\theta=0$, and (b) $\theta=\pi$ are shown in Fig. \ref{fig4}, spanning from small to large transferred momenta. To improve the clarity in the key region,  detailed low-momentum region are shown for (c) $\theta=0$ (the blue square region in (a)), (d) $\theta=0.2\pi$, (e) $\theta=0.5\pi$, and (f) $\theta=\pi$ (the green square region in (b)), corresponding to the parameters in Fig. \ref{fig3}\textcolor{blue}{(c)-(f)}.
For different $\theta$, $S(\bf{q},\omega)$ displays an anisotropic behavior arising from the non-zero COM momentum ${\bf Q}$ in an FF superfluid. At small $q$, a $\delta$-like sharp peak appears, which is the characteristic of the phonon mode originating from the spontaneously $U(1)$ symmetry breaking of pairing gap. This phonon mode starts from ${q}=0$ and exhibits linear dispersion ($\omega \propto q$) in the low-momentum region. The slope of the phonon mode at $q \rightarrow 0$ defines the sound speed, $c_{\rm s}=\omega/q$. Notably, $c_{\rm s}$ at $\theta=0$ is significantly greater than that at $\theta=\pi$. Moreover, the single-particle excitations display an angular anisotropy of $\theta$. The subsequent subsections systematically shed light on the angular anisotropy of both the collective modes and the single-particle excitations, respectively.
\subsection{Phonon mode}
\begin{figure}[h!]
\includegraphics[scale=0.4]{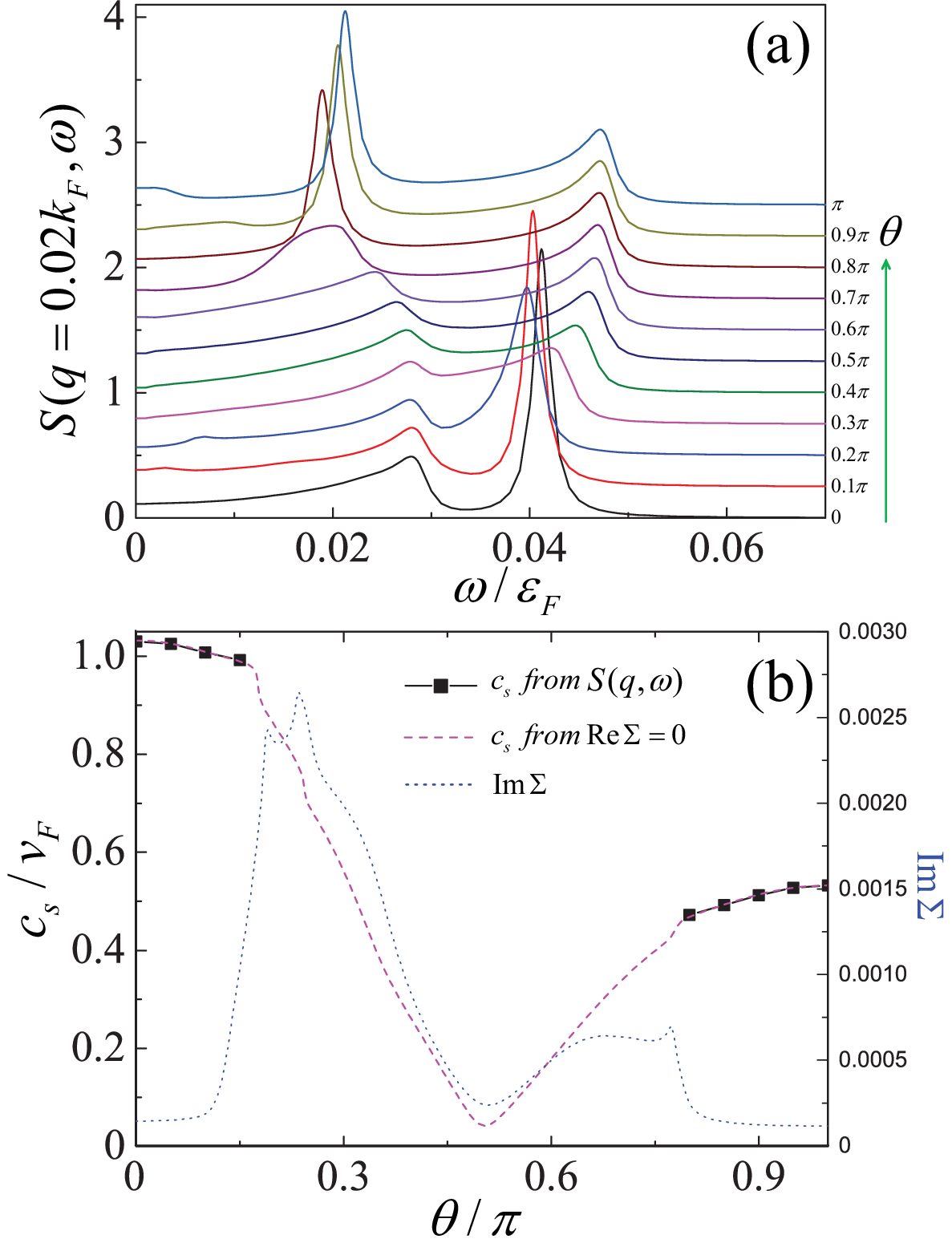}
\caption{ (a) $S({\bf q}=0.02k_{F},{\omega})$ as a function of $\omega$ for $\theta$ ranging from $0$ (bottom) to $\pi$ (top); (b) $c_{\rm s}$ versus $\theta$ with $h=0.47{\varepsilon_{F}}$, $E_{\rm b}=0.2{\varepsilon_{F}}$. The magenta dashed line denotes $c_{\rm s}$ from ${\rm Re}\Sigma=0$, while the blue dotted line responds to ${\rm Im}\Sigma$-related quasiparticle lifetime.\label{fig5}}
\end{figure}

To clearly elucidate the angular anisotropy of the phonon mode, we calculate the sound speed $c_{\rm s}$ as a function of $\theta$. The sound speed is obtained by fitting the position of the $\delta$-like peak of $S({\bf q},{\omega})$ at a small transferred momentum ($q=0.02k_{F}$). In Fig. \ref{fig5}\textcolor{blue}{(a)}, we plot the $S({q=0.02k_{F}},{\omega})$ as a function of $\omega$ for $\theta\in[0, \pi]$ from bottom to top. The extracted $c_{\rm s}$ is shown in Fig. \ref{fig5}\textcolor{blue}{(b)}.
Obviously, the dynamical excitations exhibit significant angular dependence. A sharp phonon peak emerges around $\theta=0$ and $\theta=\pi$, but vanishes for $\theta\in[0.2\pi,0.7\pi]$ due to the intense competition between the phonon mode and the single-particle excitations. Therefore, the sound speed in this angular range can not be obtained through $S({\bf q},{\omega})$. We focus on the sound speed near $\theta=0$ and $\theta=\pi$. First, the sound speed at $\theta=0$ ($c_{\rm s}=1.019v_{F}$) significantly exceeds that at $\theta=\pi$ ($c_{\rm s}=0.531v_{F}$). Second, $c_{\rm s}$ decreases with increasing $\theta$ near $\theta=0$, but it exhibits a growing trend around near $\theta=\pi$. Third, in the vicinity of $\theta=0$, the phonon mode is completely separated from the single-particle excitations, which is closely related to the opening of the band-gap $\Delta_{\rm band}$ (see Fig. \ref{fig3}). This separation vanishes when $\Delta_{\rm band}$ closes at $\theta=0.2\pi$, where the phonon mode merges into the single-particle excitation continuum and undergoes spectral broadening through the scattering with the single-particle excitations. The magenta dashed line represents the sound speed obtained through self-consistent calculation while the blue dotted line reflects the quasiparticle lifetime (discussed in Sec. \ref{discussion}).

\begin{figure}[h!]
\includegraphics[scale=0.4]{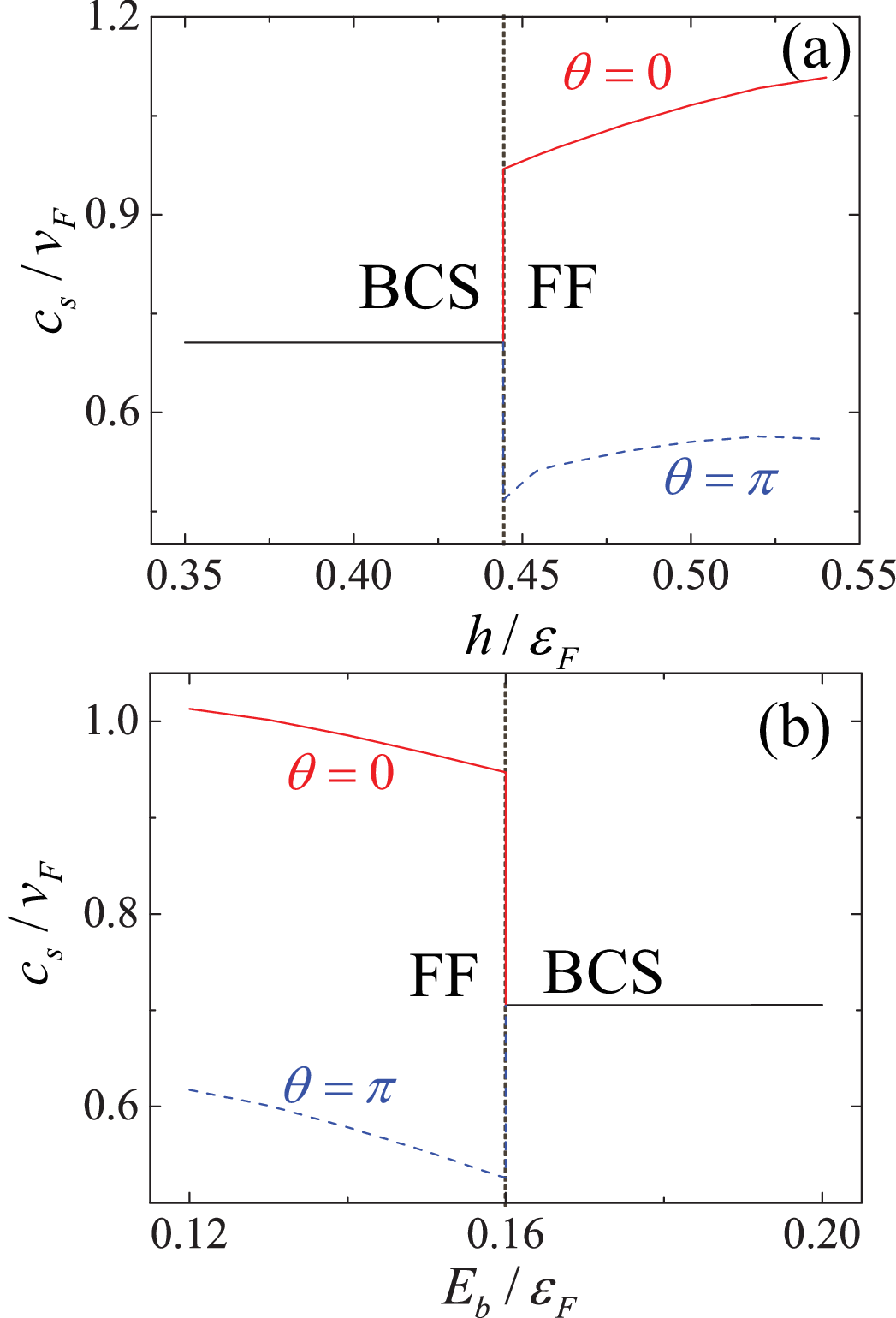}
\caption{ (a) Sound speed $c_{\rm s}$ as a function of $h$ with $E_{\rm b}=0.2{\varepsilon_{F}}$; (b) $c_{\rm s}$ versus $E_{\rm b}$ with $h=0.4{\varepsilon_{F}}$. The red solid and blue dashed lines corresponding to $\theta=0$ and $\theta=\pi$, respectively. The gray dotted lines mark the phase transition point.
\label{fig6}}
\end{figure}

The phase diagram reveals that both the Zeeman field $h$ and interaction strength $E_{\rm b}$ determine the emergence and physical properties of an FF superfluid. We systematically calculate the $h$ and $E_{\rm b}$ dependencies of the sound speed $c_{\rm s}$, as shown in Fig. \ref{fig6}.
It is shown that $c_{\rm s}$ remains almost constant with increasing $h$ in the BCS superfluid, which can be understood by the Meissner effect. When $h>h_{\rm c}$, the system comes into an FF superfluid state, where $c_{\rm s}$ abrupt increases (for $\theta=0$) or decreases (for $\theta=\pi$) and is proportional to $h$.
In an FF superfluid, the Fermi surface is reconstructed. As the quasiparticle spectrum crosses the Fermi energy, the Fermi velocity associated with $c_{\rm s}$ becomes sensitive to $h$ \cite{Belkhir1994}.
Moreover, the interaction strength can alter the quasiparticle spectra in an FF superfluid, leading to that $c_{\rm s}$ decreases with increasing $E_{\rm b}$ and has a discontinuous decrease (for $\theta=0$) or increase (for $\theta=\pi$) at the FF-BCS phase transition.

\subsection{Single-particle excitations and roton-like mode}
 Owing to the reconstruction of the Fermi surface, the single-particle excitations in an FF superfluid exhibits more complex structures compared with those in the BCS superfluid. The Cooper pair-breaking mechanism can can manifest through the single-particle excitations. As mentioned above, the two quasiparticle spectra $E_{1\bf{k}}$, $E_{2\bf{k}}$ generate four kinds of pair-breaking mechanism, namely, the intra-band excitations $\{11\}\rightarrow E^{(1)}_{\bf{k+q}}-E^{(1)}_{\bf{k}}$, $\{22\}\rightarrow E^{(2)}_{\bf{k+q}}-E^{(2)}_{\bf{k}}$ and the inter-band excitations $\{12\}\rightarrow E^{(1)}_{\bf{k+q}}+E^{(2)}_{\bf{k}}$, $\{21\}\rightarrow E^{(2)}_{\bf{k+q}}+E^{(1)}_{\bf{k}}$. Notably, the minima of intra-band excitations remain gapless while the inter-band excitations are gapped.

\begin{figure}[h!]
\includegraphics[scale=0.45]{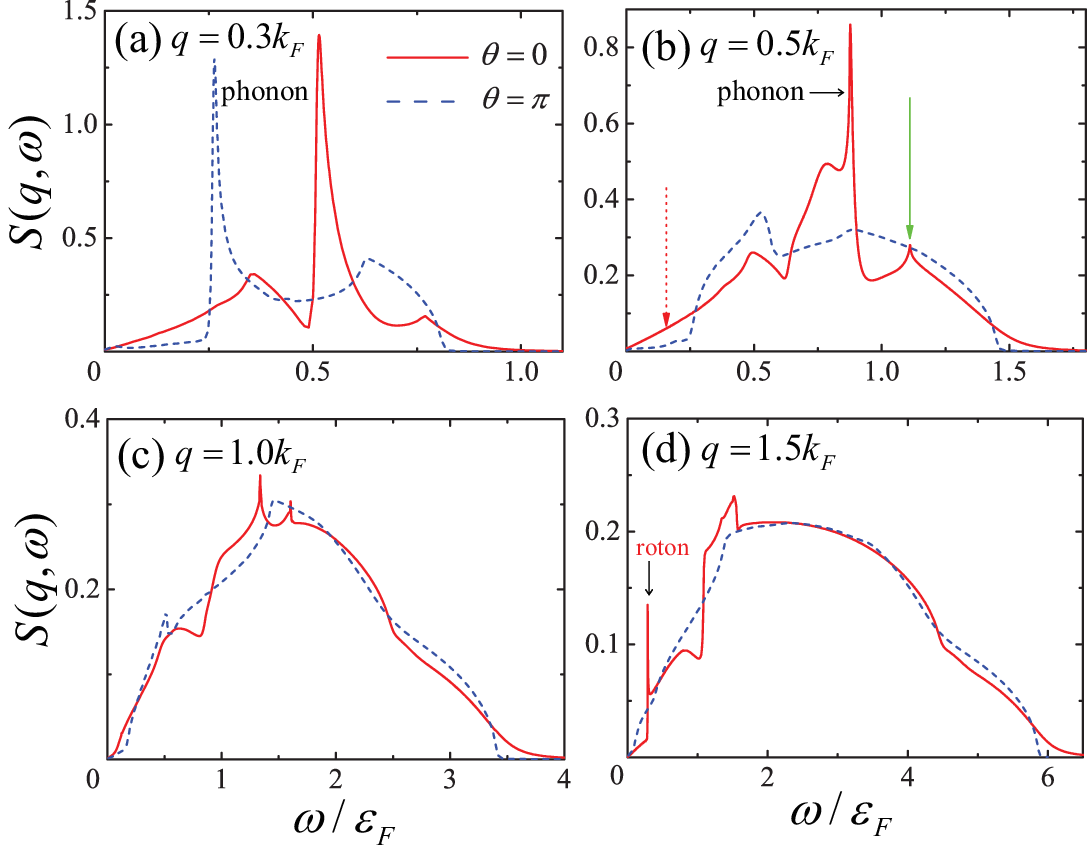}
\caption{ $S({\bf q},{\omega})$ as a function of $\omega$ for momenta (a) $q=0.3k_{F}$, (b) $q=0.5k_{F}$, (c) $q=1.0k_{F}$, and (d) $q=1.5k_{F}$ at $\theta=0$ (red solid line) and $\theta=\pi$ (blue dashed line), with $h=0.47{\varepsilon_{F}}$, $E_{\rm b}=0.2{\varepsilon_{F}}$. For $\theta=0$ in (b), the red dashed arrow indicates intra-band excitations while the green solid arrow denotes inter-band excitations. The black arrow in (d) marks the sharp peak of the roton-like mode.
\label{fig7}}
\end{figure}

 To clearly show the dynamical excitations, we calculate the energy-dependent $S({\bf q},{\omega})$ at representative transferred momenta. The results for $\theta=0$ (red solid line) and $\theta=\pi$ (blue dashed line) are plotted in Fig. \ref{fig7}.
Our calculation results demonstrate the angular anisotropy of the single-particle excitations in an FF superfluid. At $q=0.3k_{F}$, the phonon peak ($\omega=0.515{\varepsilon_{F}}$) exhibits much larger width than that at $q=0.02k_{F}$ (see Fig. \ref{fig5}\textcolor{blue}{(a)}), resulting from  the competition between the phonon mode and the single-particle excitations as the phonon mode enters the single-particle excitation continuum.
With $q$ further increases, the phonon mode gradually vanishes.
Moreover, the composition of single-particle excitations is studied. For $\theta=0$ (red line) at $q=0.5k_{F}$ in Fig. \ref{fig7}\textcolor{blue}{(b)}, the intra-band excitations dominate the nearly linear low-energy region (red dashed arrow) while the inter-band excitations determine the characteristic high-energy peak (green solid arrow).
Interestingly, at $q=1.5k_{F}$ in Fig. \ref{fig7}\textcolor{blue}{(d)}, a sharp roton-like peak located at $\omega=0.2862\varepsilon_{F}$ emerges. This roton-like mode dispersion is marked in Fig. \ref{fig4}\textcolor{blue}{(a)} (the red dashed line). The roton-like mode exhibits the angular anisotropy and disappears at a greater $\theta$.
\section{Discussion}
\label{discussion}
\begin{figure}[h!]
\includegraphics[scale=0.3]{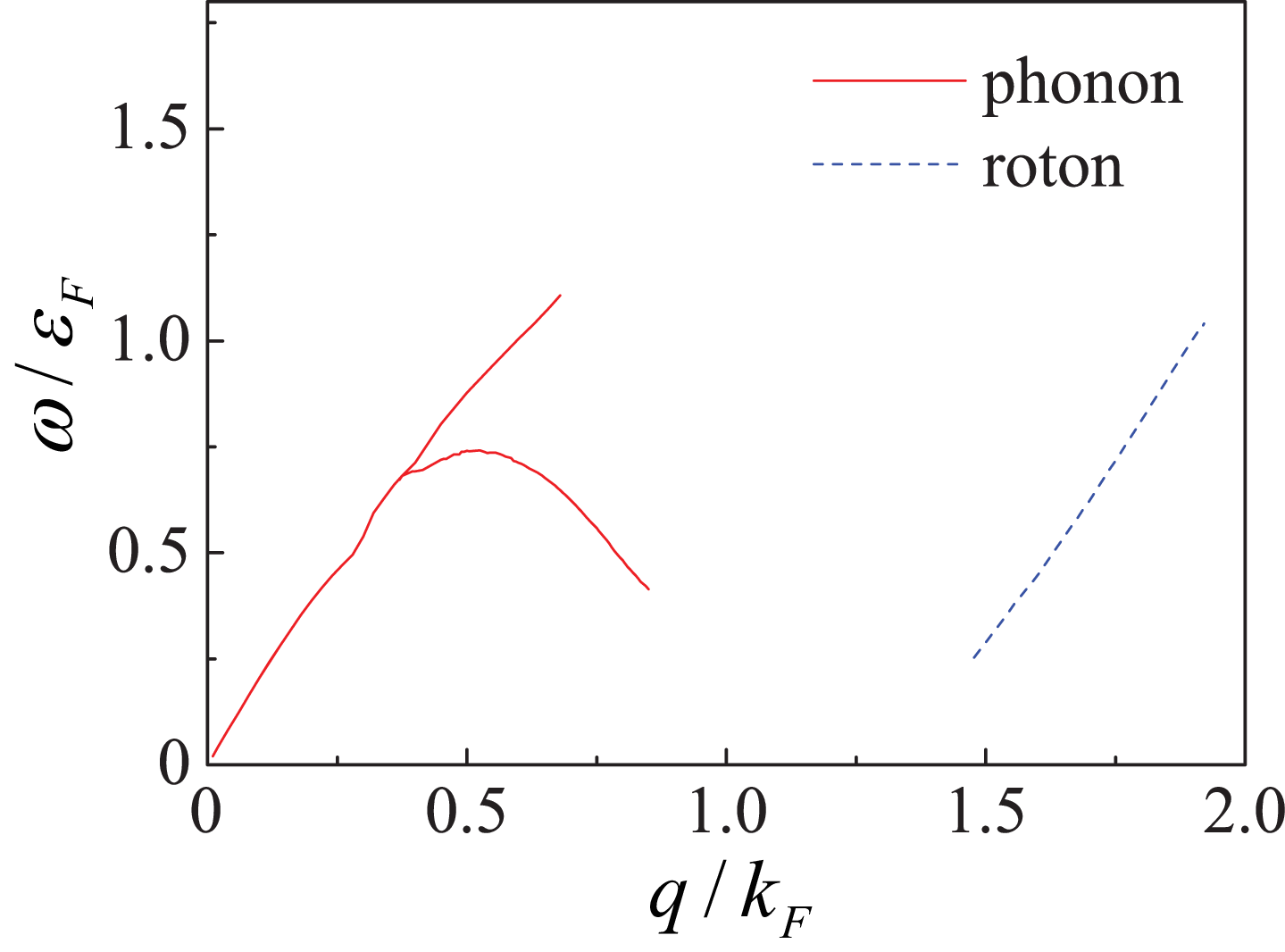}
\caption{ Solution of ${\rm Re}\Sigma=0$ as a function of $q$. The parameters are the same as those in Fig. \ref{fig4}\textcolor{blue}{(a)}.
\label{fig8}}
\end{figure}

We now try to analyze the collective modes through the RPA formula. From Eq. \ref{chi}, the collective modes corresponds to the poles of $(\hat{1}-\chi^{0}({\bf q},{i}\omega_{n})U{M_{I}})$, which can be simplified as $\Sigma=[\chi^{0}_{{\Delta}\Delta^{*}}\chi^{0}_{\Delta^{*}{\Delta}}-(\chi^{0}_{{\uparrow}{\downarrow}})^{2}]$. Following our earlier method  \cite{Zhao2020}, the dispersion of main collective modes is determined by solving the self-consistent equation, ${\rm Re}\Sigma=0$. The real part governs the quasiparticle dispersion and the imaginary part ${\rm Im}\Sigma$ determines their lifetime. In Fig. \ref{fig8}, we plot the solution of ${\rm Re}\Sigma=0$.
At small $q$, the linear dispersion (red line) corresponds to the phonon mode. As $q$ increases, the nonlinear self-consistent equation generates two sets of solutions, accounting for the emergence of the multi-branch structure in the excitation spectrum in Fig. \ref{fig4}\textcolor{blue}{(a)}. A roton-like mode emerges at large $q$. This mode is sufficiently narrow to be difficult to detect in Fig. \ref{fig4}\textcolor{blue}{(a)}, but can be clearly resolved in Fig. \ref{fig7}\textcolor{blue}{(d)}. The imaginary part ${\rm Im}\Sigma \propto 1/\tau$ ( $\tau$: quasiparticle lifetime). This inverse proportionality implies that the experimentally signal is suppressed when ${\rm Im}\Sigma$ ($\tau$) is large (small). Figure \ref{fig5}\textcolor{blue}{(b)} demonstrates that the sound speed $c_{\rm s}$ obtained from ${\rm Re}\Sigma=0$ agrees with the results from the dynamical structure factor. However, between $\theta=0.2\pi$ and $\theta=0.7\pi$, the sound speed can not be obtained through the dynamical structure factor due to the large ${\rm Im}\Sigma$.
\section{Summary}
\label{summary}
In conclusion, we propose the optimal parameters for realizing an FF superfluid in 2D polarized Fermi gases by calculating the pairing gap along the BCS-FF phase transition point. Our theoretical results reveal a dome-shaped dependence of the pairing gap at the critical point on the interaction strength, which demonstrates that excessive interaction strength suppresses the formation of an FF superfluid. Utilizing the RPA theory, we systematically study the anisotropic dynamical excitations in an FF superfluid, including the collective modes and the single-particle excitations. In an FF superfluid, the sound speed exhibits strong dependence on the Zeeman field and the interaction strength, contrasting with the BCS phase where the sound speed remains almost constant. These dynamical excitations may provide a potential scheme for identifying an FF superfluid through the two-photon Bragg scattering technique.

In the future, we will systematically study the dynamical excitations in three quantum phases: the FF superfluid on an optical lattice, the LO superfluid and the topological FF superfluid. First, compared with the continuous FF superfluid characterized by an extremely narrow parameter window, the FF states of optical lattices exhibit significantly broader parameter tunability \cite{Koponen2007}, which enhances the experimental feasibility. Moreover, we may investigate novel collective modes, including the roton mode \cite{Zhao2024,Zhang1990}. Second, in addition to the $U(1)$ symmetry breaking, an LO superfluid manifests the spontaneous translational symmetry breaking through the periodic order parameter. Thus, the collective modes of an LO superfluid are different from those of an FF superfluid, including the Higgs mode \cite{Huang2022,Fan2022}. The pairing energy of an LO superfluid is much greater than that of an FF superfluid \cite{Loh2010}. Therefore, an LO superfluid are more experimentally accessible than an FF superfluid. However, in contrast to an FF superfluid with one plane wave, the LO superfluid has two plane waves, and every atom couples to two other fermions, which makes it difficult to diagonalize the Hamiltonian \cite{Kinnunen2018,Baarsma2016}. Third, with the spin-orbit coupling interaction, the topological FF superfluids may be realized \cite{Qu2013,Xu2014,Zhang2013}. It is essential in identifying the characteristic dynamical excitations to distinguish the topological FF superfluid from the BCS superfluid.

\section{Acknowledgements}
The authors would like to thank Prof. Peng Zou and Feng Yuan for helpful discussions. This work was supported by the funds from the National
Natural Science Foundation of China under Grant No.11547034 (H.Z.), Research Foundation of Yanshan University under Grant No. 8190448 (S.T.).

Perceptually uniform color maps ('lajolla') are used in this study \cite{Crameri2018}.
\section{Appendix}
 The mean-field response function $\chi^0$ of 2D FF Fermi superfluid is numerically calculated, and all 10 independent matrices elements of $\chi^0$ are displayed as,
\begin{eqnarray}\label{a11}
\chi^{0}_{{\uparrow}{\uparrow}}&=&\sum_{\bf k}U_{\bf k+q}^2[U_{\bf k}^2I_1({\bf k},{\bf q},i\omega_n)+V_{\bf k}^2I_2({\bf k},{\bf q},i\omega_n)] \nonumber \\
&-&\sum_{\bf k}V_{\bf k+q}^2[{U_{\bf k}^2}I_3({\bf k},{\bf q},i\omega_n)-{V_{\bf k}^2}I_4({\bf k},{\bf q},i\omega_n)],\nonumber
\end{eqnarray}
\begin{eqnarray}\label{a12}
\chi^{0}_{{\uparrow}{\downarrow}}&=&-\sum_{\bf k}\frac{\Delta_0^2}{4E_{\bf k}E_{\bf k+q}}[I_5({\bf k},{\bf q},i\omega_n)-I_6({\bf k},{\bf q},i\omega_n)] \nonumber \\
&-&\sum_{\bf k}\frac{\Delta_0^2}{4E_{\bf k}E_{\bf k+q}}[I_7({\bf k},{\bf q},i\omega_n)+I_8({\bf k},{\bf q},i\omega_n)],\nonumber
\end{eqnarray}
\begin{eqnarray}\label{a13}
\chi^{0}_{{\uparrow}{\Delta}}&=&\sum_{\bf k}\frac{\Delta_0}{2E_{\bf k+q}}[{U_{\bf k}^2}I_1({\bf k},{\bf q},i\omega_n)+{V_{\bf k}^2}I_2({\bf k},{\bf q},i\omega_n)] \nonumber \\
&+&\sum_{\bf k}\frac{\Delta_0}{2E_{\bf k+q}}[{U_{\bf k}^2}I_3({\bf k},{\bf q},i\omega_n)-{V_{\bf k}^2}I_4({\bf k},{\bf q},i\omega_n)], \nonumber
\end{eqnarray}
\begin{eqnarray}\label{a14}
\chi^{0}_{{\uparrow}\Delta^{*}}&=&\sum_{\bf k}\frac{\Delta_0}{2E_{\bf k}}{U_{\bf k+q}^2}[I_1({\bf k},{\bf q},i\omega_n)-I_2({\bf k},{\bf q},i\omega_n)] \nonumber \\
&-&\sum_{\bf k}\frac{\Delta_0}{2E_{\bf k}}{V_{\bf k+q}^2}[I_3({\bf k},{\bf q},i\omega_n)+I_4({\bf k},{\bf q},i\omega_n], \nonumber
\end{eqnarray}
\begin{eqnarray}\label{a22}
\chi^{0}_{{\downarrow}{\downarrow}}&=&\sum_{\bf k}V_{\bf k}^2[{V_{\bf k+q}^2}I_5({\bf k},{\bf q},i\omega_n)+{U_{\bf k+q}^2}I_6({\bf k},{\bf q},i\omega_n)] \nonumber \\
&-&\sum_{\bf k}U_{\bf k}^2[{V_{\bf k+q}^2}I_7({\bf k},{\bf q},i\omega_n)-{U_{\bf k+q}^2}I_8({\bf k},{\bf q},i\omega_n)], \nonumber
\end{eqnarray}
\begin{eqnarray}\label{a23}
\chi^{0}_{{\downarrow}{\Delta}}&=&-\sum_{\bf k}\frac{\Delta_0}{2E_{\bf k+q}}{V_{\bf k}^2}[I_5({\bf k},{\bf q},i\omega_n)-I_6({\bf k},{\bf q},i\omega_n)] \nonumber \\
&+&\sum_{\bf k}\frac{\Delta_0}{2E_{\bf k+q}}{U_{\bf k}^2}[I_7({\bf k},{\bf q},i\omega_n)+I_8({\bf k},{\bf q},i\omega_n)], \nonumber
\end{eqnarray}
\begin{eqnarray}\label{a24}
\chi^{0}_{{\downarrow}\Delta^{*}}&=&-\sum_{\bf k}\frac{\Delta_0}{2E_{\bf k}}[{V_{\bf k+q}^2}I_5({\bf k},{\bf q},i\omega_n)+{U_{\bf k+q}^2}I_6({\bf k},{\bf q},i\omega_n)] \nonumber \\
&-&\sum_{\bf k}\frac{\Delta_0}{2E_{\bf k}}[{V_{\bf k+q}^2}I_7({\bf k},{\bf q},i\omega_n)-{U_{\bf k+q}^2}I_8({\bf k},{\bf q},i\omega_n)], \nonumber
\end{eqnarray}
\begin{eqnarray}\label{a34}
\chi^{0}_{{\Delta}\Delta^{*}}&=&\sum_{\bf k}U_{\bf k+q}^2[{V_{\bf k}^2}I_1({\bf k},{\bf q},i\omega_n)+{U_{\bf k}^2}I_2({\bf k},{\bf q},i\omega_n)] \nonumber \\
&-&\sum_{\bf k}V_{\bf k+q}^2[{V_{\bf k}^2}I_3({\bf k},{\bf q},i\omega_n)-{U_{\bf k}^2}I_4({\bf k},{\bf q},i\omega_n)], \nonumber
\end{eqnarray}
\begin{eqnarray}\label{a43}
\chi^{0}_{\Delta^{*}{\Delta}}&=&\sum_{\bf k}V_{\bf k}^2[{U_{\bf k+q}^2}I_5({\bf k},{\bf q},i\omega_n)+{V_{\bf k+q}^2}I_6({\bf k},{\bf q},i\omega_n)] \nonumber \\
&-&\sum_{\bf k}U_{\bf k}^2[{U_{\bf k+q}^2}I_7({\bf k},{\bf q},i\omega_n)-{V_{\bf k+q}^2}I_8({\bf k},{\bf q},i\omega_n)].\nonumber \\
\end{eqnarray}
The corresponding functions  $I_1({\bf k},{\bf q},i\omega_n)$, $I_2({\bf k},{\bf q},i\omega_n)$, $I_3({\bf k},{\bf q},i\omega_n)$, $I_4({\bf k},{\bf q},i\omega_n)$, $I_5({\bf k},{\bf q},i\omega_n)$, $I_6({\bf k},{\bf q},i\omega_n)$, $I_7({\bf k},{\bf q},i\omega_n)$, and $I_8({\bf k},{\bf q},i\omega_n)$ are shown as
\begin{eqnarray}\label{kenal1}
I_1({\bf k},{\bf q},i\omega_n) &=& \frac{n_F(E^{(1)}_{\bf k})-n_F(E^{(1)}_{\bf k+q})}{i\omega_n+E^{(1)}_{\bf k}-E^{(1)}_{\bf k+q}} \nonumber
\end{eqnarray}
\begin{eqnarray}\label{kenal1-1}
I_2({\bf k},{\bf q},i\omega_n) &=& \frac{1-n_F(E^{(2)}_{\bf k})-n_F(E^{(1)}_{\bf k+q})}{i\omega_n-E^{(2)}_{\bf k}-E^{(1)}_{\bf k+q}} \nonumber
\end{eqnarray}
\begin{eqnarray}\label{kenal2}
I_3({\bf k},{\bf q},i\omega_n) &=& \frac{1-n_F(E^{(1)}_{\bf k})-n_F(E^{(2)}_{\bf k+q})}{i\omega_n+E^{(1)}_{\bf k}+E^{(2)}_{\bf k+q}} \nonumber
\end{eqnarray}
\begin{eqnarray}\label{kenal2-1}
I_4({\bf k},{\bf q},i\omega_n) &=& \frac{n_F(E^{(2)}_{\bf k+q})-n_F(E^{(2)}_{\bf k})}{i\omega_n-E^{(2)}_{\bf k}+E^{(2)}_{\bf k+q}} \nonumber
\end{eqnarray}
\begin{eqnarray}\label{kenal3-1}
I_5({\bf k},{\bf q},i\omega_n) &=& \frac{n_F(E^{(1)}_{\bf Q-k-q})-n_F(E^{(1)}_{\bf Q-k})}{i\omega_n-E^{(1)}_{\bf Q-k}+E^{(1)}_{\bf Q-k-q}} \nonumber
\end{eqnarray}
\begin{eqnarray}\label{kenal3}
I_6({\bf k},{\bf q},i\omega_n) &=& \frac{1-n_F(E^{(2)}_{\bf Q-k-q})-n_F(E^{(1)}_{\bf Q-k})}{i\omega_n-E^{(1)}_{\bf Q-k}-E^{(2)}_{\bf Q-k-q}} \nonumber
\end{eqnarray}
\begin{eqnarray}\label{kenal4-1}
I_7({\bf k},{\bf q},i\omega_n) &=& \frac{1-n_F(E^{(1)}_{\bf Q-k-q})-n_F(E^{(2)}_{\bf Q-k})}{i\omega_n+E^{(2)}_{\bf Q-k}+E^{(1)}_{\bf Q-k-q}} \nonumber
\end{eqnarray}
\begin{eqnarray}\label{kenal4}
I_8({\bf k},{\bf q},i\omega_n) &=& \frac{n_F(E^{(2)}_{\bf Q-k})-n_F(E^{(2)}_{\bf Q-k-q})}{i\omega_n+E^{(2)}_{\bf Q-k}-E^{(2)}_{\bf Q-k-q}},
\end{eqnarray}
where the function $n_F(x)$ is the Fermi distribution.

\end{document}